\documentstyle[12pt]{article}
\begin{document}
\font\bss=cmr12 scaled\magstep 0
\title{\vspace{2mm}\LARGE \bf On exact solutions of
Maxwell-Bloch system for two-level medium with degeneracy}

\author{\bss S. B. Leble*,\\
\small Wydzial Fizyki
Technicznej i Matematyki Stosowanej \\
\small Politechnika 
Gdan'ska, ul. G.Narutowicza, 11/12 80-952, Gdan'sk-Wrzeszcz, Poland,\\
\small	 email leble@mifgate.pg.gda.pl\\
\bss N.V.Ustinov\\
\small Kaliningrad State University, Theoretical Physics Department,\\
\small Al.Nevsky st.,14, 236041 Kaliningrad\\
\small email ustinov.leble@theor.phys.ksu.kern.ru\\
\small *permanent adress}
\vskip 1.0true cm

\date {}
\maketitle
\begin{abstract}
Maxwell--Bloch system describing the resonant propagation of electromagnetic 
 pulses in both two--level media
 with degeneracy in angle moment projection and three--level media
with equal oscillator forces is considered.
The inhomogeneous broadening of energy levels is accounted.
Binary Darboux Transformation generating the solutions of the system
is constructed.
Pulses corresponding to the transition between levels with largest population
difference are shown to be stable.
The solution describing the propargation of pulses in the medium
exited by the periodic wave is obtained.
The hierarchy of infinitesimal symmetries is obtained by means of Darboux
transformation.

\end{abstract}

\thispagestyle{empty}
\vfill
\eject
\section{Introduction}
\mbox{}
\par
     Investigations of laser emission interaction with resonant media
sufficiently promoted the understanding and mathematical description of this
phenomena$^1$. The important kick for it gave the invention of the self-induced
 transparency (SIT) phenomena$^2$ that is a passage of a powerful and
ultrashort pulse without loss of a form and energy at a resonant medium.
The group velocity of such pulses depends on their amplitude and duration:
shorter pulse has larger velocity. It was found that while the propagation of
pulses of different duration there may be situation when a quick pulse reaches
more slow one, after "collision" both go out without change of their form and
velocities only phase shifts appear. From mathematical point of view this
wonderful stability of SIT pulses is the corollary of the complete integrability
of Maxwell-Bloch (MB) equations that describe SIT phenomenon by two-level
medium approximation
with nondegenerate levels$^3$, recently developed for the case of two-photon
processes$^4$. The SIT pulses correspond to soliton solutions of
the equations and their collision properties reflect the asymptotic properties 
of multi-
soliton states with respect to one-soliton ones.

	For real media the energy levels are usually degenerate that contribute
to peculiarities of polarizied pulses propagation. In the Ref. 5 the integrability of
MB equations for arbitrary polarization of light being resonant with quantum
transitions $j_b = 0$ $ \rightarrow $ $j_a = 1$  have been established. It was shown that
solitons with circular polarization are stable.

	This problems of interest relates to the interaction of many-frequency
laser pulses with a resonant many-level medium. For example such many-frequency
action widen possibilities of isotop dividing or simulation of chemical
reactions as well as spectroscopy investigations. The peculiarities of coherent
pulses that propagate in many-level media may be used in the frequency
transformation effectivisation$^6$.

	Let's consider a medium that may be described in the two-level atoms
approximation with energy levels $E_{a,b}$ that are degenerate by projections
m and $\mu$ of total momenta $j_{a,b}$. Let plane electromagnetic wave
propagates along z and its electric field has the form

	$${\bf E = E_o} exp[i(kz-\omega t)] + c.c. \eqno(1.1)$$
where t is time. The carrier frequency $ \omega = kc$ of the pulse is close to
$\omega_o = 2\pi (E_b - E_a)/h$ of the atom transition $j_a \rightarrow j_b$.

The evolution of the pulse (1) in the semiclassical approach and coherent
approximation is described by generalized MB equations$^5$ that may be represented
in the dimensionless form as

$$\epsilon_{q;\zeta} = \sum_{\mu,m} R_{\mu,m} <J^q_{\mu,m}>$$.

$$R_{\mu,m;\tau} = i \sum_{q,} \epsilon_q (\sum_{m'} J^q_{\mu,m'}R_{m',m} -
\sum_{\mu'} R_{\mu,\mu'} J^q_{\mu',m})$$

$$R_{m,m';\tau} = i \sum_{q,m} ( \epsilon_q^* J^q_{\mu,m} R_{\mu,m'} -
\epsilon_q R_{m,\mu} J^q_{\mu,m'})$$

$$ R_{\mu,\mu';\tau} = i \sum_{q,m} ( \epsilon_q J^q_{\mu,m} R_{m,\mu'} -
\epsilon_q^* R_{\mu,m} J^q_{\mu',m})$$

In those equations the amplitude $E_o$ of a light pulse, coefficients of the
density matrix	$ \rho_{\mu,\mu}$, are connected with the functions $\epsilon
_q, R_{\mu,\mu'}, R_{m,m'}, R_{\mu,m}$:

$$E_q = h\epsilon_q/2\pi t_od,$$

$$\rho_{\mu,\mu'} = N_o f(\eta) R_{\mu,\mu'}, $$

$$\rho_{m,m'} = N_o f(\eta) R_{m,m'},$$

$$\rho_{\mu,m} = N_o f(\eta) R_{\mu,m} exp[i(kz - \omega t)]$$
here d is the reduced dipole moment of the transition $j_b -> j_a, N_o$ -
density
of resonant particles, $t_o = (3h/2\pi \omega d ^2 N_o)^{1/2}$ is time dimension
constant. $\Delta = (\omega - \omega_o) t_o$ - resonance shift.
$\tau = (t - z/c)/t_o, \zeta = z/ct_o$,

\[J^q_{\mu,m} = (-1)^{j_b-m}
\left[
\begin{array}{ccc}
j_a &1 &j_b\\
-m &q &\mu
\end{array}
\right] 3^{1/2}\]
by the index q the spherical components of the vector ${\bf \epsilon}$ are
labelled .

	Let's consider the transition $j_b = 0 -> j_a = 1$ and introduce
$$e_+ = - i\epsilon_{+1}, e_- = -i\epsilon_{-1},$$
$$n_b = R_{\mu\mu},|_{\mu = \mu' = 0},$$
$$n_{a+}= R_{mm'}|_{m=m'=1}, n_{a-}= R_{mm'}|_{m=m'=-1}$$
$$\nu_a = R_{mm'}|_{m = m' = -1}, \nu_+ = R_{\mu m}|_{\mu = 0, m = 1},\nu_- =
R_{\mu m}|_{\mu = 0, m = -1} \eqno(1.2)$$
The quantization axis is choosen along the light propagation direction
(axis $\zeta$)

As the interaction of a quantum system with the transverse electromagnetic
waves is accompanied by transitions with angular momentum change (-1 or +1),
the MB system may be simplified.

$$e_{-,\zeta} = - <\nu_->$$
$$e_{+,\zeta} = - <\nu_+> \eqno(1.3a)$$
$$n_{a-,\tau} = - \nu_-e_-^* - \nu_-^*e_- $$
$$ n_{a+,\tau} = - \nu_+e_+^* - \nu_+^*e_+$$
$$ n_{b,\tau} = \nu_+ e_+^* - \nu_+^*e_+ + \nu_-e_-^* + \nu_-^* e_- \eqno(1.3b) $$

$$ \nu_{-,\tau} = - i(\eta - \Delta)\nu_- + (n_{a-} - n_b)e_- + \nu_a e_+$$
$$ \nu_{+,\tau} = - i(\eta - \Delta)\nu_+ + (n_{a+} - n_b)e_+ + \nu_a e_-$$
$$\nu_{a,\tau} = - \nu_+e_-^* - \nu_-^* e_+ \eqno(1,3c)$$

Here the angle brackets denote the inhomogeneous broadening averaging with
normalized partition function $f(\eta)$
$$
<F> = \int_{-\infty}^{\infty} f(\eta) F d \eta \,\,.
$$

If the resonant atoms ensemble is in pure state, one may rest only five
equations in the system (1.3).

$$e_{-\zeta} = - <a_3a_1^*> $$
$$e_{+\zeta} = - <a_3a_2^*> \eqno(1.4a)$$
$$a_{1,\tau} = i(\eta - \Delta)a_1/2 - a_3 e_-^*$$
$$a_{2,\tau} = i(\eta - \Delta)a_2/2 - a_3 e_+^*$$
$$a_{3,\tau} = i(\eta - \Delta)a_3/2 + a_1 e_- + a_2 e_+\eqno(1.4b)$$

where $a_{1,2,3}$ are amplitudes of the probability density (population numbers)
of the lower and two upper levels.

The same system of nonlinear equations describes the resonant propagation of
two--frequency radiation with equal shifts from resonance in three-level
medium where one of the transitions is forbidden and the oscillator forces of
the rest two transitions are equal.

\bigskip

\section{Zero curvature representation, reductions of
potentials and Darboux transformation}
\mbox{}
\par
The Maxwell - Bloch equations (1.2) represents a complicated system that is not
solved analytically in the case of arbitrary angular momenta. However in the
case of resonant transitions $j_b \rightarrow j_a = 1 ; j_b \rightarrow j_a
= 0$ the equations (1.2) may be presented as the compatibility condition of
some linear system (zero curvature representation) with a sperctral parameter
(SP) $\lambda$.
For the case of equations (1.3) it is the following system
$$\psi_{\tau} = V \psi$$
$$\psi_{\zeta} = <\alpha(\lambda)A>\psi \eqno (2.1)$$
where $V = U - \lambda J, J = diag\{1,1,-1\}, \alpha(\lambda) = (2\lambda +
i(\eta- \Delta))^{-1},$
\[U =\left(
 \begin{array}{ccc}
0& 0& -e^*_-\\
0& 0& -e^*_+\\
e_-& e_+& 0\\
\end{array} \right)\]

\[A =\left(
 \begin{array}{ccc}
n_{a-}& \nu_a &\nu^*_-0\\
\nu^*_a& n_{a+}& \nu^*_+\\
\nu_-& \nu_+& n_b\\
\end{array} \right)\]

From the obvious equality $\psi_{\zeta\tau} = \psi_{\zeta\tau}$ we obtain
the system
$$U_\zeta = [J,<A>]/2,$$
$$A_\tau = [U,A] - i(\eta - \Delta)[A,J]/2, \eqno(2.2)$$
from which it follows that the elements of U and A matrices should satisfy
MB system (1.3).
 If matrix elements of A are representable in the form of
$A_{kj} = a_ka_j^*  (k,j = 1,2,3)$ (the pure quantum states) then from (1.3)
the system (1.4) follows.

The equations (1.3,4) appear as the compatibility conditions of the conjugate
Lax pair as well
$$\xi_{\tau} = - \xi W$$
$$\xi_{\zeta} = - \xi <\alpha(\kappa) A> \eqno(2.3)$$
where $W = U - \kappa J$.

As the matrices U and A satisfy the following reduction

$$ \left\{
 \begin{array}{cc}
  U + U^+ = 0 \\
 A - A^+ = 0,
\end{array} \right.
\eqno(2.4)
$$
one may check that there in the space of matrix solutions of ZS equations exists
the authomorphism in the sense of the Ref. 7,
$$\Psi(\lambda) \rightarrow [\Psi(-\lambda^*)^+]^{-1}\in\Psi(\lambda).\eqno(2.4)$$
By the way there is the coupling between the WF spaces of right and conjugate
(left) zero curvature representation

$$\kappa = - \lambda^*, \xi = \psi^+\eqno (2.5)$$
Such functions are refered as coupled by an authomorphism.

The following statement is valid: The equations (1.3) are the corollary of (2.2)
when the additional condition (2.4) is posed.

The DT technique that is developed in the Refs. 8,9 may be applied for the construction
of solutions hierarchy for the MB system (1.3). We specialize the scheme here.
Let $\phi$ and $\chi$ are functions of	the right and left (conjugate) problems
(2.1,3). Spectral parameters (SP) are $\mu$ and $\nu$ correspondingly. We also
suppose that those functions are connected by the authomorphism and
$$\nu = -\mu^*, \chi = \phi^*.$$
The transformed WFs by the binar$y^8$ DT
$$\psi[1] = [1 - (\mu + \mu^*)P/(\lambda + \mu^*)]\psi \eqno(2.6a)$$
$$\xi[1] = \xi [1 - (\mu +\ mu^*)P/(\kappa - \mu)] \eqno(2.6b)$$
are also the solutions of right and conjugate ZS zero curvature reprezentations
with new potentials $V[1] = U[1] - \lambda  J$, $W[1] = U[1] - \kappa J$ and
$$A[1] = A - 2(\mu + \mu^*)[\alpha(\mu)AP + \alpha(\mu)^*PA -
(\alpha(\mu)A + \alpha(\mu)^*)PAP], \eqno(2.6c)$$
where
$$U[1] = U - (\mu + \mu^*)[J,P]. \eqno(2.6d)$$
The matrix P at the equations (2.6) is defined in analogy with Ref. 8 by
$$P_{kj} = \phi_k\phi_j^*/(\phi^+,\phi)$$
As $P^+ = P$ the reduction (2.4) is conserved. It means that elements of matrices
U[1] and A[1] give new solutions of MB equations (1.3). For the transformed
potential  coefficients that have the sense of dimensionless 
electromagnetic field components the following expressions are valid
$$e_-[1] = e_- + 2(\mu + \mu^*)\phi_3\phi^*_1/(\phi^+,\phi) \eqno(2.7a)$$
$$e_+[1] = e_+ + 2(\mu + \mu^*)\phi_3\phi^*_2/(\phi^+,\phi) \eqno(2.7b)$$
If the initial state of the quantum system was pure the matrix $A_{kj} =
a_ka_j^*$ and the result of the transformation is pure as well with
$${\bf a[1]} = {\bf a} - 2(\mu + \mu^*)\alpha(\mu)^*P{\bf a},$$
the normalization is conserved too.

As by the map (2.6) the transforms of WFs coupled by an authomorphism should also
be connected by $\xi[1] = \psi[1]^+$ the reduction (2.4) will be valid by
iterations of BDT. Final result may be expressed via $\phi^{(q)}$
 and $\chi^{(q)}$ that are WFs of initial zero curvature representation to be
coupled by
authomorphisms in pairs$^9$.
\medskip

\section {Solutions over the zero and periodic backgrounds.}
\mbox{}
\par
In this section the construction of solutions of MB (and NS) systems by the 
transformation (2.6,7) will be considered.

{\bf 3.1} Starting from zero functions $e_- = e_+ = 0, n_{a+,-} = \nu_a = 
\nu_{+,-} = 0,$
but $n_b = 1$ we obtain after first iteration the one-soliton solution that generalize
$2\pi$ pulse of two-state system. This solutions have been built in the 
paper$^5$ within the IST method. However a construction of two-soliton solution 
met difficulties and the analysis of soliton collisions was made by asympotic 
methods$^6$. By the way in the Ref. 10 the algebraic approach that have features 
of B{\"a}cklund and "Dressing" methods was developed and the breather 
two-soliton 
(with equal  velocities) solution have been found.

After a second iteration preserving the reduction 
(the WF have SP coupled by $\mu_{(2)} = \mu_{(1)^*}$) 
we obtain the solution of MB system with potential elements

$$e_-[2] = -2(a_1c_1^*\Delta_2\exp[- \imath\eta] - a_2c_2^*\Delta_1\exp[\imath\eta] -
a_1c_2^*\Delta^* - a_2c_1^*\Delta)/\Delta[2] $$
$$e_+[2] = - 2(b_1c_1^*\Delta_2\exp[- \imath\eta] - b_2c_2^*\Delta_1\exp[\imath\eta] -
b_1c_2^*\Delta^* - b_2c_1^*\Delta)/\Delta[2] $$
$$ \Delta[2] = \Delta_1\Delta_2 - \mid\Delta\mid^2$$
$$\Delta_1 = [(\mid a_1\mid^2 + \mid b_1\mid^2)\exp[-\vartheta] +
\mid c_1\mid^2\exp[\vartheta]]/(2\mu_R)$$
$$\Delta_2 = [(\mid a_2\mid^2 + \mid b_2\mid^2)\exp[-\vartheta] +
\mid c_2\mid^2\exp[\vartheta]]/(2\mu_R)$$
$$\Delta = [(a_1 a^*_2 + b_1 b^*_2) \exp[ - \vartheta - \imath theta] + 
c_1c^*_2 \exp[\vartheta + \imath \theta]]/(2\mu)$$
where
$$\vartheta = 2\mu_R(\tau + <\mid\alpha(\mu)\mid^2>\xi),$$
$$\theta = 2 \mu_I \tau - <(2\mu_I + \eta - \Delta)\mid \alpha(\mu) \mid^2>\xi),$$
where $\mu = \mu_R + \imath \mu_I, a_1{,2}, b{1,2}, c_{1,2}$ are complex constants.
The transformation of such a structure we may also name the binary one$^{1,8}.$
This solution generalize the solution from Ref. 10. It is characterized by eight
real parameters: $\mu_R,\mu_I$, distance between pulse centers, electric field
components ratio of the pulses and phase shifts of the same component of both
pulses. At Ref. 10 the solution contain four parameters with fixed amplitude and
phase shift ratios.

{\bf 3.2} Let us consider the case of arbitrary level populations 
("nonzero dipole temperature"
of the medium): $n_{a+}n_{a-}n_b \neq 0, e_- = e_+ = \nu_- = \nu_+ = \nu_a = 0$.
WFs of the sistem (2.1) have the form
$$\psi_1 = C_1 exp ( - \lambda \tau + <\alpha(\lambda)n_{a-}>\zeta)$$
$$\psi_2 = C_2 exp ( - \lambda \tau + <\alpha(\lambda)n_{a+}>\zeta)$$
$$\psi_3 = C_3 exp ( \lambda \tau + <\alpha(\lambda)n_b>\zeta)$$
Transforming by (2.6) with WF $\phi$ and spectral parameter $\mu$ one get
$$e_-[1] = 4\mu_R\phi_3\phi_1^*/(\phi^+,\phi),$$
$$e_+[1] = 4\mu_R\phi_3\phi_2^*/(\phi^+,\phi),$$
where
$$\phi_1 = C_1\exp( - \mu\tau + <\alpha(\mu)(n_{a-} - n_{a+})>\zeta/2)$$
$$\phi_2 = C_2\exp( - \mu\tau + <\alpha(\mu)(n_{a+} - n_{a-})>\zeta/2)$$ 
$$\phi_3 = C_3\exp( \mu\tau + <\alpha(\mu)(2n_b- n_{a-} - n_{a+})>\zeta/2)$$

Let $n_b > n_{a+} > n_{a-}$. Then while $\tau \rightarrow - \infty,  e_+[1]
\rightarrow 0 $ at arbitrary $\zeta$ and if $\tau \rightarrow + \infty$, then
$e_-[1] \rightarrow 0$. So the solution describes a transformation of
a pulse resonant with the
transition which has the minor population difference to one that corresponds
to the larger
population difference. It means that such pulses are not stable. In the
paper $^6$ that process the more cumbersome IST technique in terms of Riemann -
Hilbert problem we see the opposite statement . Our result seems to be more
physical: the transition from the basic state to more populated one would be
more transparent.

{\bf 3.3} Now we shall construct solutions of the MB system starting from a periodic
background.  Let 
the plane electromagnetic wave is
propagated along one of transitions 

$$e_+ = E\exp\imath(k\zeta + \omega\tau), \qquad e_- = 0,$$
where $E$ is the a complex constant.
As in MB equations the shift from resonance
is taken into account we may put $\omega = 0$.
Let also the second level is
empty $n_{a-} = 0, n_b + n_{a+} = 1$. Then from the  equations (1.3) one have

$$\nu_- = \nu_a = 0$$,
$$\nu_+ = \imath(2n_b - 1)(\omega + \eta - \Delta)^{-1} E \exp\imath(k\zeta +
\omega\tau),$$
$$k = - < (2n_b - 1)(\omega + \eta - \Delta)^{-1}>,$$
$$n_b = (1 \bar+ (\omega + \eta - \Delta) ( 4\mid E \mid^2 + (\omega + \eta -
\Delta)^{-1/2} )/2. $$
During the derivation of the lower level population $n_b$ it was supposed
that the quantum system had gone in the considered state from one with free
upper levels. This condition fix the trace of the matrix $A^2$ that does not
depend on $\tau$. It may be seen from the second equation of the system (2.2).
The case of the arbitrary higher level population may be considered
analoguesly. Relaxation terms from spontaneous transitions may be neglected
if the dimensionless electric field amplitude $\mid E\mid $ is big enough.
The solutions of the linear system are:

$$ \psi_1 = C_1 exp( - \lambda\tau),$$
$$ \psi_2 = (C_+ exp(\vartheta) + C_- exp( -\vartheta)) exp((<\alpha(\lambda)> -
ik) \zeta/2) $$
$$ \psi_3 = - (C_+ (\lambda+\sigma))exp(\vartheta) + C_- (\lambda-\sigma)
exp( -\vartheta)) exp((<\alpha(\lambda)> +
ik) \zeta/2)/E^*,$$
where
$$\vartheta = \sigma (\tau + \imath <(2n_b - 1)\alpha(\lambda)(\eta - 
\Delta)^{-1}>\zeta)$$
$$\sigma = (\lambda^2 - \mid E \mid^2)^-{1/2}.$$
Transforming by (2.6) and introducing a new parameter $\gamma = \gamma_R +
i \gamma_I$ instead of the SP $\mu$ by 
$$ \cosh(\gamma) = \mu / \mid E \mid $$
one obtains
$$e_-[1] = - 4 E \cosh(\gamma_R) \cos (\gamma_I) \bar{\phi_3} \bar{\phi}_1
\exp(ik\zeta/2)/(\bar{\phi}^+,\bar{\phi}),$$
$$e_+[1] = -  E (1-
4Ecosh(\gamma_R) cos (\gamma_I) \bar{\phi_3} \bar{\phi}_2)
exp(ik\zeta)/(\bar{\phi}^+ \bar{\phi}),$$
where
$$\bar{\phi}_1 = C_1 \exp( - \mid E \mid \cosh(\gamma_R) \cos(\gamma_I)(\tau +
 <D>\zeta) +
$$
$$
\imath (\mid E \mid \sinh(\gamma_R) \sin(\gamma_I) \tau) - <D(\eta -
\Delta)> \zeta/2)).$$
$$ \bar{\phi}_2 = C_+ \exp( \vartheta) + C_- \exp ( - \vartheta ).$$
$$ \bar{\phi}_3 = -E( C_+ \exp( \vartheta + \gamma) + C_- \exp ( - \vartheta -
\gamma ))/\mid E \mid,$$
and $\vartheta = Re \vartheta + \imath Im \vartheta,$
$$ Re \vartheta = \mid E \mid \cos(\gamma_I)(\sinh(\gamma_R) \tau +
$$
$$<(2n_b - 1)(\eta - \Delta)^{-1} ((\eta - \Delta) \sinh(\gamma_R) -
 2 \mid E \mid sin (\gamma_I)) D> \zeta),
$$
$$  Im \vartheta = \mid E \mid \cosh(\gamma_R)(\sin(\gamma_I) \tau +
<(2n_b - 1)(\eta - \Delta)^{-1} ((\eta - \Delta) \sin(\gamma_I) +
$$
$$
+ 2 \mid E \mid \sinh (\gamma_R)) D> \zeta),
$$
$$ D = 2 \mid E \mid^2 (\cosh (2 \gamma_R) + \cos(2 \gamma_I)) + (\eta - \Delta)^2
$$
This solution is parametrized by four complex parameters $E,
 \gamma, C_1, C_+/C_-$ . 
The obtained solutions describe
proceses of fission and fusion of radiation pulses that are resonant with
different transitions. It shows the existence
of pulses with
different propagation velocities over the periodic wave background. The
propagation of SIT pulses at arbitrary level poulation also demonstrate
processes of frequency transformation. 
In order to verify this
it is necessary to consider real parts of exponents indices in the expressions
for $\bar\phi_1,
\bar\phi_2,
 \bar\phi_3$.
The parameter $\gamma$ is introduced namely for a convenience of this operation.
The expressions that determine the character of pulses behaviours are however
complicated but allow to plot electric field components. It is seen that the
amplitude $ e_+$ tends to $\mid E \mid$ at large $\zeta$ and the phase of the
periodic wave undergo a shift.

The technique developed here may
be applied for many-level systems in the cases when the corresponding
MB equations are integrable. The  studying of more complicated reductions
of the ZS problem potencials seems to be interesting
looking for more weak constraints on quantum systems parameters. Another
possibility for applications is investigation of the conditions of pulse
propagation in thin
films
as well as relection from them.

\section {Infinitesimal parameters of DT.}
\par
In this section the solution for a linearization of the equations
(2.2) to be obtained . The dependence of this solution
on an arbitrary function give the possibility to widen classes 
of
solutions of initial - boundary problems, looking for sequence of infinitesimal
symmetries $^{11}$ and for the stability of soliton solutions analisys.
The main observation
that allows to develop the technique, is that the nontrivial binary transformations (2.6)
may generate a new potential that coincide with the seed one $^{12}$. 
In this article BDT (not Matveev one) and the method of keeping reduction
were used.  
It means that the DT is infinitesimal if
$U[1] = U$, if $\nu = \mu$ and $(\chi, \,\varphi) \ne 0$.
\par
Linearizing the system (1.7) with respect to perturbations
$U^{(1)}$ and $A^{(1)}$
in relations to solutions $U$ and  $A$ we get
$$
\left\{
\begin{array}{l}
U^{(1)}_{\zeta} = \displaystyle\frac{1}{2}
[ \, J\, , \, < \! A^{(1)} \! > \, ] \\
A^{(1)}_{\tau} = \displaystyle \left[ \, \frac{i}{2}(\eta - \Delta)J +
U \, , \, A^{(1)} \, \right] + [ \, U^{(1)} \, , A \, ]
\end{array}
\right. .
\eqno (4.1)
$$
We shall suppose that  $\nu = \mu + \delta$,
and $\delta$ is small .
\par
Expanding (2.6) in Taylor series by the parameter $\delta$ we get
$$
\left\{
\begin{array}{l}
U[1] = U + \delta \, [ \, J \, , \, P^{(0)} \, ] + ... \\
A[1] = A + 2\delta \alpha(\mu) \, [ \, A \, , \, P^{(0)} \, ] + ...
\end{array}
\right. ,
$$
where $P^{(0)} = \varphi \otimes \chi$\,, $\varphi$ and $\chi$ are solutions
of the direct and conjugate problems (2.8,9) with the spectral parameter
$\mu$, such that
$(\chi, \varphi) = 1$.
\par
From these results the following formulas for small disturbances are derived
$$
\left\{
\begin{array}{l}
U^{(1)} = [ \, J \, , \, P^{(0)} \, ]  \\
A^{(1)} = 2\alpha(\mu) \, [ \, A \, , \, P^{(0)} \, ]
\end{array}
\right. .
\eqno (4.2)
$$
\par
After iterations of binary DT with the conditions
$$
\nu^{(q)} = \mu^{(q)} + \delta^{(q)} \,\, ,
$$
where
$$
\delta^{(q)} = O\left(\delta^{(1)}\right) \,\,\, (q = \overline{1,N})
$$
and
$$
|\delta^{(1)}| << |\mu^{(1)}| \,\, ,
$$
one get
$$
\left\{
\begin{array}{l}
U^{(1)} = \displaystyle \left[ \, J \, , \,
\sum^N_{i=1} \beta_i \varphi^{(i)} \chi^{(i)} \, \right]  \\
A^{(1)} = \displaystyle 2 \left[ \, A \, , \,
\sum^N_{i=1} \alpha(\mu^{(i)}) \alpha_i \varphi^{(i)}
\chi^{(i)} \, \right]
\end{array}
\right. ,
$$
where $\beta_i$ --- are constants.
\par
Going from sums to integrals we obtain the following
representations of the solutions
of the system (2.2)
$$
\left\{
\begin{array}{l}
U^{(1)} = \displaystyle \left[ \, J \, , \,
\int_C \Psi(\lambda) \Phi(\lambda) d\lambda \, \right]	\\
A^{(1)} = \displaystyle 2 \left[ \, A \, , \, \int_C \alpha(\lambda)
\Psi(\lambda) \Phi(\lambda) d\lambda \, \right]
\end{array}
\right. ,
\eqno (4.3)
$$
where $\Psi (\lambda)$ and $\Phi (\lambda)$ --- are matrix solutions of the basic
and conjugate problems (2.1,3) with a spectral parameter $\lambda$.
\par
The validity of the formulas (4,2,3) one may check by the direct
substitution in the equation (4.1).
\par
We notice also that the reduction (2.5) is compatible with the equation (3.3).
By means of equality (3.5) it is possible to obtain solutions that
satisfy the reduction, if integrals in the right-hand-side
would be presented as a sum by contour integrals that may be parametrized
in the following way
$$
\displaystyle \int_C \Psi(\lambda) \Phi(\lambda) d\lambda =
\int_{t_1}^{t_2} \Psi(g(t)) \Phi(g(t)) g'(t) dt +
$$
$$
\displaystyle + \int_{t_1}^{t_2} \Phi^+(g(t)) \Psi^+(g(t)) g^*\,'(t) dt \,\, ,
$$
$$
\displaystyle \int_C \alpha(\lambda) \Psi(\lambda) \Phi(\lambda) d\lambda =
\int_{t_1}^{t_2} \alpha (g(t)) \Psi(g(t)) \Phi(g(t)) g'(t) dt -
$$
$$
\displaystyle - \int_{t_1}^{t_2} \alpha^*(g(t)) \Phi^+(g(t))
\Psi^+(g(t)) g^*\,'(t) dt \,\, .
$$
\par
The formulas of this paragraph are novel ones.
As these expressions for small perturbations contain arbitrary matrix
solutions of both problems (2.1,3) they may be applied for studying
of stability of soliton solutions of the MB equation.
It is also possible to construct infinitesimal symmetries $^{11,12}$ of the
equations . and widen classes of initial - boundary problems solutions as well.

\section{Conclusion}
\par
We would repeat that in $^{12}$ the infinitesimal binary DT was introduced 
and exploited to extract hierarchies of symmetries
The technique of DT with reductions described above may also be applied to the 
case of NSS (Manakov equation)$^{13}$ as the first equation and the reduction of 
the potential are 
the same. The only difference is in formulas (2.6a,b) and (2.7): the vector WF 
should be replaced by solutions of the consistent Lax problem$^{12}$.
Analoguesly one can built solutions of the NSS over the periodic background
(the soliton case is considered at Ref. 6). Let

$$e_- = 0, \qquad e_+ = E \exp (i(k \zeta + \omega \tau))$$
Then from Lax equations it follows that
$k = \omega^2 - \mid E \mid^2. $
 Solving the Lax system one get expressions for WF
$$ \psi_1 = C_1 exp( - \lambda\tau + i\lambda^2 \zeta),$$
$$ \psi_2 = (C_+ exp(\vartheta) + C_- exp( -\vartheta)) exp(- i(k\zeta +
\omega\tau0)/2) $$
$$ \psi_3 = - (C_+ (\lambda+\sigma - ik/2)) exp(\vartheta) + C_- (\lambda-
\sigma - ik/2) exp( -\vartheta)) exp(i(k \zeta + \omega \tau),$$
where
$$
\vartheta = \sigma ( \tau + i ( \lambda + ik/2) \zeta)
$$
$$
\sigma = (( \lambda - ik/2)^2 - \mid E \mid^2)^-1/2.
$$
If you put here $\lambda = \mu$ and substitute the obtained formula for
$ \phi $ in the equation (2.7) you'll get the desired solution. Its
difference from the solution of MB system is only in other dependence on
the variable $\zeta$.

It is possible to obtain the integrable equation that combine systems
MB and NS. This new integrable system may describe a propagation of polarized
waves in Kerr nonlinear media with resonant atoms admixture similarily to 
linear polarized case of Ref. 14.
The similar approach to perturbation theory based on Riemann - Hilbert problem
is presented at the recent paper $^{15}$.

\section{Acknoledgement} 

One of us (S. Leble) thanks H.Steudel and N.Sasa for discussions and 
some useful references information.

\section{Referencies}

$^1$
R.K.Boullough, P.M.Jack, P.W.Kitchenside and R.Saunders, Physica  Scripta ,
 {\bf20}, 364 (1979).

$^2$
S.L. McCall, E.L. Hahn, Phys. Rev. Lett., {\bf18}1, 908 (1967).

$^3$
G.L. Lamb, Rev.Mod.Phys. {\bf 43}, 99 (1979).

$^4$
H. Steudel, D.J. Kaup
J.Mod.Opt. {\bf 43}, 1851 (1966).
H. Steudel, R. Meinel, D.J. Kaup . Solutions of degenerate two-photon propagation
from B{\"a}cklund transformations, J.Mod.Phys., to appear.

$^5$
A.M. Basharov , A.I. Maimistov 
. JETP {\bf 87}, 1595 (1984).

$^6$
L.A. Bolshov , N.N. Elkin, V.V. Likhansky, M.I.Persiantsev 
. JETP {\bf 94} 101 (1988).

$^7$
A. Mikhailov 
The reduction problem and the inverse scattering method. Physica D {\bf3}
73 (1981).

$^8$
Leble S.B., Ustinov N.V.Solitons of Nonlinear Equations Associated with
Degenerate Spectral Problem of the Third Order.In: International Symposium
on Nonlinear Theory and its Applications (NOLTA 93) Hawaii, U.S.A.,
December 5-10, 1993, 4.8-1,p.547-550.

$^9$
S.B. Leble , N.V. Ustinov .:
Deep reductions for matrix Lax system,
invariant forms and elementary Darboux transforms. in {\it Proceeding 
of NEEDS-92
Workshop, World Scientific, Singapore} (1993), p.34-41. J. Phys. A: Math.Gen.
{\bf26} 5007 (1993).

$^{10}$
H. Steudel in {\it Proceedings of 3d International Workshop on Nonlinear
Processes in Physics. Eds. V.G. Bar'yakhtar et al. Kiev, Naukova Dumka} (1988)
Vol.1, p. 144.

$^{11}$
N.R. Sibgatullin Dokl. AN SSSR, {\bf 291} p.302 (1986).

$^{12}$ Ustinov N.V. ''The reduced self--dual Yang--Mills equation, binary and
infinitesimal Darboux Transformations.'' Journal of Mathematical
Physics, , {\bf39}, pp.976-985 (1998).

$^{13}$
S.V. Manakov 
JETP {\bf 65},  505 (1973).

$^{14}$ S. Kakei J. Satsuma Multi-Soliton Solutions of a Coupled System
 of the Nonlinear Schr\"odinger Equation
and the Maxwell-Bloch Equations, to be published.

$^{15}$
V.S. Shchesnovich,  Chaos, Solitons and Fractals {\bf 5} p.2121-2133 (1997).

\end{document}